\documentclass[12pt,a4paper,final]{iopart}

\usepackage{iopams}  
\usepackage[breaklinks=true,colorlinks=true,linkcolor=blue,urlcolor=blue,citecolor=blue]{hyperref}
\usepackage[dvips]{graphicx}
\usepackage{bm}

\usepackage[latin9]{inputenc}
\usepackage{textcomp}

\newcommand{\rxz}{\mathrm{xz}}
\newcommand{\ryz}{\mathrm{yz}}
\newcommand{\rz}{\mathrm{3z^2-r^2}}

\begin{document}
\title[Conductance through $3d$ impurities in gold chains]{Kondo behavior and conductance 
through $3d$ impurities in gold chains doped with oxygen}
\author{M. A. Barral, S. Di Napoli}
\address{Dpto de F\'{\i}sica de la Materia Condensada, GIyA-CNEA, Avenida General Paz 1499, 1650 San Mart\'{\i}n,
Provincia de Buenos Aires, Argentina, CONICET, 1033 CABA, Argentina}
\author{G. Blesio}
\address{Instituto de F\'{\i}sica Rosario. Facultad de Ciencias Exactas, Ingenier\'{\i}a y Agrimensura, 
Universidad Nacional de Rosario, CONICET, Bv. 27 de Febrero 210 bis, 2000 Rosario, Argentina}
\author{P. Roura-Bas}
\address{Dpto de F\'{\i}sica de la Materia Condensada, GIyA-CNEA, Avenida General Paz 1499, 1650 San Mart\'{\i}n,
Provincia de Buenos Aires, Argentina, CONICET, 1033 CABA, Argentina}
\author{Alberto Camjayi}
\address{Departamento de F\'{\i}sica, FCEyN, Universidad de Buenos Aires and IFIBA, 
Pabell\'on I, Ciudad Universitaria, CONICET, 1428 CABA, Argentina} 
\author{L. O. Manuel}
\address{Instituto de F\'{\i}sica Rosario. Facultad de Ciencias Exactas, Ingenier\'{\i}a y Agrimensura, 
Universidad Nacional de Rosario, CONICET, Bv. 27 de Febrero 210 bis, 2000 Rosario, Argentina}
\author{A. A. Aligia}
\address{Centro At\'{o}mico Bariloche and Instituto Balseiro, Comisi\'{o}n Nacional
de Energ\'{\i}a At\'{o}mica, CONICET, 8400 Bariloche, Argentina}
\date{\today }

\begin{abstract}
Combining {\it ab initio} calculations and effective models derived from them,
we discuss the electronic structure of oxygen doped gold chains when one Au atom
is replaced by any transition-metal atom of the $3d$ series. The effect of O doping
is to bring extended Au $5d_{xz}$ and $5d_{yz}$ states to the Fermi level, which together with the
Au states of zero angular momentum projection, lead to three possible channels for the screening 
of the magnetism of the impurity.
For most $3d$ impurities the expected physics is similar to that of the underscreened Kondo model,
with singular Fermi liquid behavior. For Fe and Co under a tetragonal crystal field
introduced by leads, the system might display non-Fermi liquid behavior. 
Ni and Cu impurities are described by a $S=1$ two channel Kondo model and an SU(4) impurity Anderson model
in the intermediate valence regime, respectively. 
In both cases, the system is a Fermi liquid, but the conductance shows some observable differences  
with the ordinary SU(2) Anderson model.
\end{abstract}



\maketitle


\section{Introduction} 
\label{intro}

In the last years there has been a great amount of experiments in electronic transport through semiconducting 
\cite{gold1, cro, gold2, wiel, grobis, ama, keller} and molecular 
\cite{kuba,parks,parks2,serge,mina} quantum dots (QDs), in which 
manifestations of the Kondo effect were observed.
In addition, experiments with mechanically controllable break
junctions made possible to create one-dimensional atomic chains of several elements \cite{ohni,smit,rodri},
and the conductance of noble-metal nanowires with transition-metals impurities has been measured 
using point-contact techniques \cite{eno,bak}.

The Kondo effect is one of the paradigms in strongly correlated
condensed matter systems  \cite{hew} and arises when the spin of a localized electron 
(such as that of a magnetic $3d$ impurity) is at least partially screened 
by conduction electrons interacting with the spin. In its simplest version, a 
localized spin 1/2 and the conduction electrons form a 
many-body singlet ground state.  
The binding energy of this singlet is of the order of the characteristic Kondo temperature 
$T_K$ below which the effects of the ``screening'' of the impurity spin manifest themselves in different physical 
properties, such as the conductance as a function of temperature $G(T)$, which has been measured 
in quantum dots and is in excellent agreement
with theory \cite{grobis}.

In the general case, the localized spin has magnitude $S$ and can be screened by $N$ channels,
which correspond to conduction states of different symmetry. The case $2S=N$ (including 
the simplest one $2S=N=1$) corresponds to the {\em fully screened} case with a singlet ground state.
For $2S>N$ one has the {\em underscreened} case, in which the total spin of the ground state is 
$S-N/2$ and the scattering of conduction electrons near the Fermi energy 
corresponds to {\em singular} Fermi liquid behavior \cite{meh,logans}.
Finally $2S<N$ corresponds to the {\em overscreened} Kondo effect, in which 
inelastic scattering persists even at vanishing temperatures and excitation
energies, leading to a non-Fermi liquid. 
The simplest overscreened model is the two-channel Kondo one, $S=1/2$, $N=2$, 
in which the impurity contribution to the entropy is $\ln(2)/2$ and the conductance per
channel at low $T$ has the form $G(T)\simeq G_{0}/2\pm a\sqrt{T}$,
where $G_{0}$ is the conductance at zero temperature in the one-channel
case \cite{zar,mit}.

The underscreened Kondo effect and
quantum phase transitions involving partially Kondo screened spin-1
molecular states were observed in transport experiments by changing
externally controlled parameters \cite{parks2,serge}.
On the other hand, a system consisting in a Co atom inside a Au chain, in which the symmetry
is reduced to four-fold by connecting it to appropriate leads has been proposed
as a possible realization of the overscreened Kondo model \cite{dinap,dinap2,dinap3}.
Here, the role of the two conduction channels is played by the $5d_{xz}$ and $5d_{yz}$ electrons of Au, 
which are equivalent by symmetry (asymmetry between the channels destroys the non-Fermi liquid 
behavior \cite{mit}).

We note that for one localized electron with two conduction channels equivalent by symmetry 
(for example $xz$ and $yz$ in tetragonal symmetry), one expect orbital SU(2) symmetry, 
which combined with the spin SU(2) one, leads to SU(2) $\times$ SU(2) symmetry.
However often in practice the symmetry corresponds to a larger Lie group, the SU(4) one, 
leading to another exotic Kondo effect. In fact the low-energy effective Hamiltonian for a molecule of
iron(II) phtalocyanine on Au(111) has SU(4) symmetry \cite{mina,joaq}.
This is also the case for quantum dots in carbon nanotubes  \cite{jari,choi,lim,ander,lipi,buss,fcm,grove,see}. 
Recently, the SU(4) Kondo effect has been argued to correspond to experimental
observations of a system of two quantum dots for a  certain range of parameters 
\cite{keller,oks,nishi,fili,bao,restor,restor2}. As we show below, a Cu impurity
in a Au chain doped with O can be described by the SU(4) Anderson model, which can be 
thought as a generalization of the Kondo model to include charge fluctuations.
If instead of one, we have two localized electrons, the Coulomb repulsion breaks the
SU(4) symmetry. If in the system above, the Cu impurity is replaced by Ni, 
the system can be described by a spin 1, two channel Kondo model 
with SU(2) $\times$ SU(2) symmetry \cite{nickel}, which is also analyzed in this work.

In this paper, we study the electronic structure of $3d$ impurities in oxygen doped gold chains.
Using {\it ab initio} calculations we determine the occupancy of the different spin-orbitals at the 
impurity site. 
In some problematic cases, we also use continuous-time quantum Monte Carlo with three orbitals to solve 
the impurity, in a similar way as in dynamical mean-field theory (DMFT).
From this information, we infer the effective Anderson or Kondo model that 
describes the system at low energies.
The case for a Co impurity \cite{dinap,dinap2,dinap3} and a Ni one \cite{nickel} were studied
before. 
Using the numerical 
renormalization-group (NRG) applied to the effective model, we calculate explicitly the conductance 
as a function of temperature for 
the Ni impurity showing that it has a behavior qualitatively similar as for ordinary one-channel 
$S=1/2$ Kondo impurities but  with important and observable quantitative differences.
We also calculate the conductance for a Cu impurity, corresponding to the SU(4) Anderson model.

In section  \ref{methods} we explain the different methods used in this work. 
Section \ref{res} contains the results for the whole $3d$ series.
In Section \ref{summ} we summarize the results.

\section{Methods}
\label{methods}

The basis of this study consists in {\it ab initio} calculations based on density
functional theory (DFT). We use the full potential linearized
augmented plane waves method, as implemented in the WIEN2K code~\cite{wien2k}.
The generalized gradient approximation for the exchange and correlation potential in the parametrization of 
PBE (Perdew, Burke and Ernzerhof)~\cite{PBE96} and the augmented plane waves local orbital basis are used. The cutoff parameter which gives the 
number of plane waves in the interstitial region  is taken as $R_{MT}*K_{max} = 7$, where $K_{max}$ is the value of the 
largest reciprocal lattice vector used in the plane waves expansion and $R_{MT}$ is the smallest muffin tin radius 
used. The number of \textbf{k} points in the Brillouin zone is enough, in each case, to obtain the desired energy  and 
charge precisions, namely 10$^{-4}$ Ry and 10$^{-4}$e, respectively. In all the studied cases we consider a  
periodically repeated hexagonal lattice with $a=b=15$~bohr and  the coordinate system is fixed in such a way that the 
chain axis is aligned with  $z$. The $a$ and $b$ distances in the supercell were checked to be large enough to avoid 
artificial interactions between the periodic replicas of the wire.

As it is known, transport experiments in Au chains indicate that the conduction channel is a single $6s$ 
band~\cite{Rego03, Ruitenbeek03}. However, due to self-interaction errors, DFT calculations can yield a spurious 
magnetization in a Au wire  at its equilibrium distance, $d_{Au-Au}^{eq}=4.9285$~bohr~\cite{Tosatti08, Tosatti09, 
DalCorso13}. In order to avoid it, we include a Hubbard $U=4eV$ correction in the $5d$-electron Au manifold, as 
discussed in Ref.~{\cite{nickel}}.

As mentioned in Section~\ref{intro}, the purpose of this work is to study the Kondo physics in cases  where 
the localized spin has magnitude $S$ and can be screened by $N$ channels, which correspond to conduction states of 
different symmetry. One possible experimental realization for obtaining the $N$ channels in Au wires 
is the incorporation of O-dopants, which make the $5d_{xz,yz}$ orbitals of Au cross the Fermi level, 
due to the large hybridization with the oxygen $2p_{x,y}$ states and the transfer of electrons to them
due to the larger electronegativity of oxygen. In our previous work~\cite{nickel}, we 
have determined that an O-doping of 19\% is the minimal amount needed in order to open the $\left|m\right|=1$~-symmetry conduction channel 
through all the Au atoms in the chain. The Au atoms that have an O atom as a nearest neighbor become slightly spin
polarized. To avoid a magnetic interaction with the localized spin, in most of our calculations these Au atoms
are placed away from the impurity.

\begin{figure}[ht]
\begin{center}
\includegraphics[width=10cm]{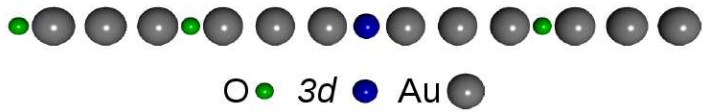}
\caption{(Color online) Schematic representation of the unit cell used for the 
$3d$ impurity embedded in a 19\% O-doped 
Au chain.}
\label{3d-imp}
\end{center}
\end{figure}

In Fig.~\ref{3d-imp} we show a schematic representation of the 16-atom unit cell used in our \textit{ab initio} 
calculations. The 3$d$  impurity is placed in the center of the O-doped Au chain and it has two O atoms symmetrically 
located as fourth neighbors. In this configuration the effect of O-doping is the desired one and does 
not alter neither the spin-state nor the 
symmetries that each 3$d$ impurity has when embedded in a bare Au 
chain.
To set the distances between atoms in O-doped Au chains, 
we relax the Au-O distance for the case of AuO diatomic chain (two-atom unit cell) and, 
afterwards,  we take the same bond length, $ d_{Au-O}^ {eq}=3.625$~bohr, for all the studied chains. 
In a similar way, 
we relax all the 3$d$ impurity-Au distances taking into account the corresponding 
two-atom unit cell and fix the obtained 
length in the subsequent calculations.

Due to the presence of localized 3$d$ electrons in a low dimensional system, we test the effects of electron 
correlations in the orbital occupancies, by including a variable Hubbard U parameter in the transition metal 3$d$ 
orbitals. Among the different possibilities for the GGA+U approach, we use the self-interaction correction
variant~\cite{Anisimov93}. For some impurities, the spin state as well as the hole/electron's symmetry are stable when 
taking into account the correlations. However, there are a few cases (Cr and Fe) where GGA and GGA+U give qualitatively different 
results. In order to elucidate the charge distribution for those cases, we use the continuous-time quantum Monte Carlo 
(CTQMC) \cite{cont1,cont2}. For simplicity in these calculations we include only those orbitals which have a 
significant hybridization between the impurity and the rest of the chain, namely those with angular momentum projection 
$m = \pm 1$ and $m=0$. In all the treated cases, the states with $m = \pm 2$ are the most localized, as they lie 
perpendicular to the chain, thus having a small hybridization.

Restricting the Coulomb interaction to these states \cite{kroll}, the effective Hamiltonian for the impurity 
takes the form

\begin{eqnarray}
&H_I = U \sum_\alpha n_{\alpha,\uparrow} n_{\alpha,\downarrow} + J_\mathrm{t} \sum_{\sigma_1,\sigma_2} d^\dagger_{\rxz,\sigma_1}d^\dagger_{\ryz,\sigma_2}d_{\rxz,\sigma_2}d_{\ryz,\sigma_1} + J_\mathrm{t}\sum_{\beta \neq \gamma}d^\dagger_{\beta,\uparrow}d^\dagger_{\beta,\downarrow}d_{\gamma,\downarrow}d_{\gamma,\uparrow} \nonumber \\
&+ (U - 2J_\mathrm{t})\sum_{\sigma_1,\sigma_2} n_{\rxz,\sigma_1} n_{\ryz,\sigma_2} 
+ (U - 2J_\mathrm{b})\sum_\zeta \sum_{\sigma_1,\sigma_2} n_{\rz,\sigma_1} n_{\zeta,\sigma_2} \nonumber \\
&+ J_\mathrm{b} \sum_\zeta \sum_{\sigma_1,\sigma_2}d^\dagger_{\rz,\sigma_1}d^\dagger_{\zeta,\sigma_2}d_{\rz,\sigma_2}d_{\zeta,\sigma_1} \nonumber \\
&+ J_\mathrm{b} \sum_\zeta \left( d^\dagger_{\rz,\uparrow}d^\dagger_{\rz,\downarrow}d_{\zeta,\downarrow}d_{\zeta,\uparrow} + \mathrm{H. c.} \right), \label{eq:imp} 
\end{eqnarray}
where $d^\dagger_{\alpha,\sigma}$ ($d_{\alpha,\sigma}$) are creation (destruction) operators, 
$n_{\alpha,\sigma}$ the  
number operator, and the sum with $\alpha$ index runs over all orbitals ($3z^2-r^2$, $xz$, $yz$) 
while the sums with other Greek 
index run over degenerate $xz$, $yz$ symmetries.
For an isolated impurity, we assume that the on-site energy of all $3d$ orbitals is the same. 
This neglects part of the crystal field due to interatomic Coulomb interactions, but this is usually
much smaller than the effect of hopping with the neighbors which is included in our treatment \cite{kroll}. 
For all cases we fix $U=4$~eV, $J_\mathrm{t}=0.70$~eV and $J_\mathrm{b}=0.49$~eV. 
The last two values result from a fit of the lowest atomic-energy levels of the $3d$ series.
For a detailed discussion of the general values of these parameters see Ref. \cite{kroll}.

The hybridization of the impurity with the Au-chains can be taken into account defining an hybridization matrix 
$\Delta$ which is simply related to the Green's function of the extreme site of the chain, close to the impurity. 
In Matsubara frequencies $i \omega_n = (2n -1) \pi T$, with $T$ the temperature in eV, it reads
\begin{equation}
\left[\Delta(i \omega_n) \right]_{\alpha,\alpha} = t_\alpha^2 \left[G^0_L(i \omega_n) + G^0_R(i \omega_n) 
\right]_{\alpha,\alpha}.\label{eq:delta}
\end{equation}
Here $G^0_{L(R)}$ are the Green's function of the nearest site to the left (right) of the impurity. 
They were obtained by first solving an isolated O-doped Au-chain with an empty site instead of a 
$3d$-impurity, by using GGA and then calculating the Green's function 
by definition, using the local density of states. The hopping parameters $t_\alpha$ were estimated 
from the fact that the imaginary part of the hybridization 
at zero
frequency (after performing an analytic continuation), 
$\Gamma_\alpha = -\mathcal{I}m\Delta(0)_{\alpha,\alpha}=\pi t_\alpha^2(\rho_L(0)+\rho_R(0))$, can be related with 
the half-width
of the peak corresponding to the GGA $3d$-impurity density of states. For the studied cases of Cr and Fe we obtain  
$t_{3z^2-r^2}=0.38$ eV, while $t_{xz,yz}=0.7$ eV for Cr and $t_{xz,yz}=1.05$ eV for Fe. 
It might seem surprising that $t_{3z^2-r^2} < t_{xz,yz}$, since the hopping between two $d$ orbitals with $m=0$
is expected to be larger than the corresponding hopping for $m = \pm 1$. However, the relevant parameter 
is $\Gamma_\alpha$ and it is always larger for $\alpha=3z^2-r^2$.

The impurity (\ref{eq:imp}) and its hybridization with the chains (\ref{eq:delta}), define an effective
multiorbital Anderson model that can be solved by continuous time quantum Monte Carlo. Here we use an hybridization
 expansion based algorithm \cite{cont1,cont3}.

As we will show in Sec.~\ref{res}, for the Cr impurity we fix its total occupancy to two electrons while for Fe we considered two cases with total 
occupancies of three and four electrons. 

To calculate the conductance in specific cases, we use the numerical renormalization group 
\cite{nrg-1, nrg-2} as implemented in the NRG Ljubljana open source code \cite{nrg-3,nrg-4}. We take the discretization 
parameter $\Lambda = 3$ and we keep up to 10000 states.

\section{Results}
\label{res}

In Table~\ref{tab1} we present the {\it ab initio} results for the orbital projected electronic occupancies 
obtained integrating the charge in each 
muffin tin, for the first five members of the $3d$ series, when considering $U=0$ at the impurity sites. The effect 
of $U$ is discussed below. 
We note that the {\it ab initio} methods discussed in the previous sections are mean-field approximations and
predict a long-range ferromagnetic order, which is not expected in a one dimensional system. This has been proved for 
example for the periodic Anderson model.\cite{noce} In spite of this, we believe that the predicted charge 
distribution for the magnetic impurity is in general reliable for a given orientation of the spin of the Hund rules
ground state and consistent with CTQMC, which does not break SU(2) symmetry. 

\begin{table}[ht!]
\caption{Orbital projected electronic occupancies for majority spin for the different elements. All the minority spin 
$d$ orbitals are almost empty.}
\label{tab1}
\begin{indented}
\item[]\begin{tabular}{@{}cccc}
\br
3$d$ impurity   &$n(3d_{3z^2-r^2}\uparrow)$& $n(3d_{xy,x^2-y^2}\uparrow)$ &  $n(3d_{xz,yz}\uparrow)$\\
\mr
Sc  & 0.12       	   & 0.31             		  & 0.13         	               \\
Ti  & 0.43       	   & 0.99              		  & 0.16          	               \\
V   & 0.77       	   & 1.64            		  & 0.21         	               \\
Cr  & 0.80       	   & 1.72           		  & 1.23        	               \\
Mn  & 0.91       	   & 1.86               	  & 1.86                                \\
\br
\end{tabular}
\end{indented}
\end{table}

From the resulting occupancies we expect the following behavior for the different 
impurities:

\subsection*{Sc}
This case should correspond to only one electron at the impurity site. The obtained partial occupancies indicate that 
this electron occupies one of the degenerate orbitals $d_{x^2-y^2}$, $d_{xy}$. Since the top of the Au states with the 
same symmetry lies about 1.3 eV below the Fermi energy, the states with one of these orbitals occupied have practically 
no hybridization with the Au states. Thus one expects a localized spin 1/2 and no Kondo effect. The mixing with other 
configurations suggested by the occupancies of Table \ref{tab1} are probably an artifact of the GGA. 

\subsection*{Ti}
One electron occupies one of the degenerate orbitals $d_{x^2-y^2}$, $d_{xy}$ and another one 
partially occupies the $d_{3z^2-r^2}$ orbital. One can describe the system as valence fluctuations between a 
configuration with one ``frozen'' electron (because it has no hopping) with $|m|= 2$ and another one
with two electrons (the other electron occupying the $d_{3z^2-r^2}$ orbital) with spin 1 due to Hund's rules. This model was 
solved exactly by Bethe ansatz \cite{bet1,bet2,bet3,bet4}. The physics corresponds to that of the underscreened $S=1$, 
one-channel Kondo model.

\subsection*{V}
Two electrons occupy each of the inert orbitals $d_{x^2-y^2}$, $d_{xy}$ with majority spin. A third electron
fluctuates between the localized $d_{3z^2-r^2}$ orbital at V and the Au band with the same symmetry. The occupancy of the V 
$3d_{3z^2-r^2}$ orbital is high suggesting that the model that describes the system is the $S=3/2$ one-channel Kondo one.
If charge fluctuations are important, the model that describes the system is the one that mixes the
$S=3/2$ configuration with the two-electron $S=1$ one. This model is also exactly solvable \cite{bet2,bet3,bet4}.
The ground state has total spin 1 and therefore corresponds to the underscreened case.

\subsection*{Cr}
As for V, the inert orbitals with majority spin are occupied. In addition, it seems that there is a little bit 
less than one electron in the $d_{3z^2-r^2}$ orbital and a little bit more than one electron in the degenerate
orbitals $d_{xz}$ and $d_{yz}$. Assuming integer occupancies one has an orbitally degenerate  
$S=2$ impurity screened by three channels. One would expect a two-stage partial screening, first by the 
$5d_{3z^2-r^2}$ Au conduction states which have a larger hybridization, and then by the $5d_{xz}$ and $5d_{yz}$,
with some similarities to iron(II) phtalocyanine molecules on Au(111) \cite{mina}, but here the ground state 
would have $S=1$.

\begin{figure}[ht!]
\begin{center}
\includegraphics[width=0.49\columnwidth]{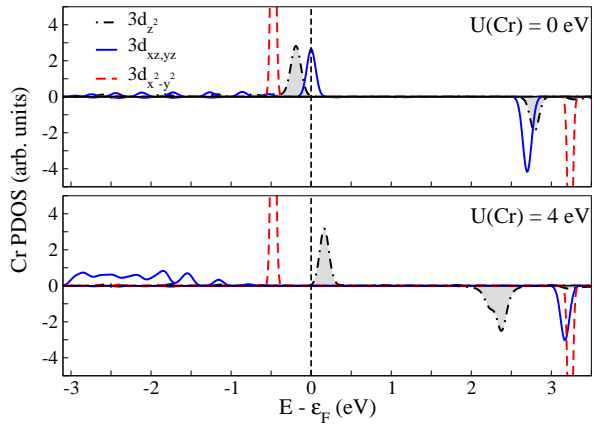} 
\caption{(Color online) Partial density of states for the Cr impurity embedded in an O-doped gold chain with 
$U=0$ (top) and $U=4$~eV (bottom) at the impurity site.}
\label{ldau1}
\end{center}
\end{figure}

\begin{table}[ht!]
\caption{Orbital projected electronic occupancies of the Cr majority spin for different values of $U$ at the Cr site.}
\label{tabu1}
\begin{indented}
\item[]\begin{tabular}{@{}cccc}
\br
$U$ (eV) &$n(3d_{3z^2-r^2}\uparrow)$& $n(3d_{xy,x^2-y^2}\uparrow)$ &  $n(3d_{xz,yz}\uparrow)$ \\
\mr
\textit{0} & 0.80       	     & 1.72               	      & 1.23          	 \\
\textit{2} & 0.82       	     & 1.73              	      & 1.27        	 \\
\textit{3} & 0.20       	     & 1.76                           & 1.75        	 \\
\textit{4} & 0.15       	     & 1.77           		      & 1.77        	 \\
\br
\end{tabular} 
\end{indented}
\end{table}

In order to study in more detail the charge distribution between the hybridized orbitals with $|m|<2$, we have performed 
GGA+U calculations introducing the interaction $U$ in the $3d$ shell. We have to mention that in the case of V no 
appreciable changes are seen when $U$ is introduced. For this reason we omit the details for V. Instead, from 
Fig.~\ref{ldau1}, we see that the effect of $U$ for Cr is to push up the $3d_{3z^2-r^2}$ orbital and transfer the electron 
to the $|m|=1$ orbitals. This is more clearly seen in Table \ref{tabu1} where the partial occupancies are shown.
The results for $U=4$~eV suggest a ground state with all orbitals with $|m|>0$ occupied forming a $S=2$ orbitally 
non-degenerate ground state screened by two degenerate channels ($m= \pm 1$).

In attempt to clarify the discrepancies in the charge distribution as a function of $U$ in the 
GGA+U calculation, we have used CTQMC as outlined in the previous section. We assume that one electron occupies 
each of the inert orbitals $m= \pm 2$, with parallel spins, and adjust the chemical potential in such a way that two
additional electrons are present in the system formed by the other orbitals with $|m| <2$, the interactions among which 
is described by Eq. (\ref{ham}). The resulting charge distribution indicates that these 
two additional electrons occupying each of the $|m|=1$ orbitals. 
This distribution agrees with GGA+U for large $U$.
In spite of the complexity introduced by the presence of two channels, We expect a similar physics to the 
one-channel $S>1/2$ underscreened Kondo model, with singular Fermi liquid behavior.

\subsection*{Mn}
As expected, this case corresponds to an $S=5/2$ Kondo model, partially screened by the three conduction channels 
($|m|<2$) at low enough temperatures.

In Table~\ref{tab2} we list the orbital projected electronic occupancies for the remaining members of the $3d$ series, 
except for Zn, which is expected to have a full $3d$ shell and therefore not showing interesting physics. In all the 
listed cases we consider $U=0$ at the impurity sites.

\begin{table}[ht!]
\caption{Orbital projected electronic occupancies for minority spin for the different elements. All the majority spin 
$d$ orbitals are occupied.}
\label{tab2}
\begin{indented}
\item[]\begin{tabular}{@{}cccc}
\br
Case	     &$n(3d_{3z^2-r^2}\downarrow)$& $n(3d_{xy,x^2-y^2}\downarrow)$ &  $n(3d_{xz,yz}\downarrow)$ \\
\mr
\textit{Fe}  & 0.19       	     & 0.60               	      & 0.39          	           \\
\textit{Co}  & 0.88       	     & 1.17              	      & 0.22        	           \\
\textit{Ni}  & 0.92       	     & 1.87             	      & 0.56         	           \\
\textit{Cu}  & 0.83       	     & 1.89           		      & 1.87        	           \\
\br
\end{tabular} 
\end{indented}
\end{table}

\subsection*{Fe}
It has six electrons, or four holes in the $3d$ shell. 
The GGA results point to 1.40 holes in the inert $|m|=2$ orbitals, 0.8 holes in the 
$m=0$ one and 1.61 holes in the $|m|=1$ orbitals.  However, since the $|m|=2$ electrons do not hop, 
one expects an integer population of them. If the occupancy 
is 1 and the remaining 3 holes occupy each of the remaining orbitals ($d_{3z^2-r^2}$, $d_{xz}$, $d_{xz}$),
the configuration is similar to that of Co below except for the 
presence of the $3d_{3z^2-r^2}$ hole. Interestingly after a first-stage Kondo effect in which the 
spin of this orbital is 
screened (as in iron(II) phtalocyanine on Au(111)~\cite{mina}), the physics would be the same as 
that for a Co impurity 
(discussed below).

\begin{figure}[ht!]
\begin{center}
\includegraphics[width=0.49\columnwidth]{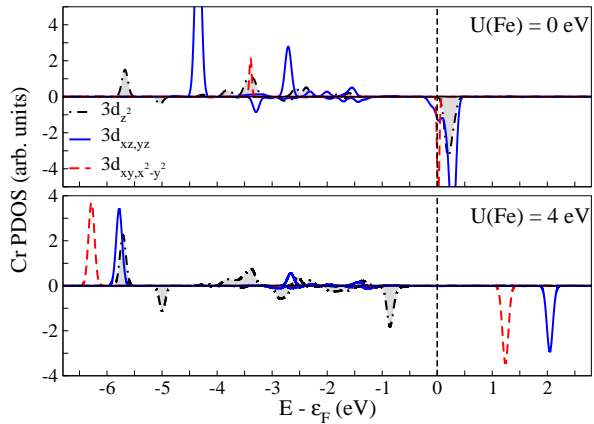} 
\caption{(Color online) Partial density of states for the Fe impurity embedded in an O-doped gold chain with 
$U=0$ (top) and $U=4$~eV (bottom) at the impurity site.}
\label{ldau2}
\end{center}
\end{figure}

In this case, we have also investigated the effect of $U$ within GGA+U to gain more insight into the charge 
distribution.
Taking into account the results shown in Fig.~\ref{ldau2}  and the occupancies listed in Table \ref{tabu2},    
we see here that the effect of $U$ is to push down the $d_{3z^2-r^2}$ orbital, increasing abruptly its occupancy between
$U=2$ eV and $U=3$ eV. In this case, all holes are in $3d$ states with $|m|>0$. By an electron-hole transformation, 
the physics is the same as that suggested by GGA+U for large $U$ for Cr: an $S=2$ orbitally non-degenerate ground state 
screened by two degenerate channels.

\begin{table*}[ht!]
\caption{Orbital projected electronic occupancies of the Fe minority spin for different values of $U$ at the Fe site.}
\label{tabu2}
\begin{indented}
\item[]\begin{tabular}{@{}cccc}
\br
U(Fe) (eV) &$n(3d_{3z^2-r^2}\downarrow)$& $n(3d_{xy,x^2-y^2}\downarrow)$ &  $n(3d_{xz,yz}\downarrow)$ \\
\mr
\textit{0} & 0.19       	     & 0.60               	      & 0.39          	         \\
\textit{2} & 0.16       	     & 0.69              	      & 0.23        	         \\
\textit{3} & 0.89       	     & 0.01                           & 0.13        	         \\
\textit{4} & 0.89       	     & 0.01           		      & 0.11        	         \\
\br
\end{tabular}
\end{indented}
\end{table*}

As for the case of Cr, we have studied the charge distribution with CTQMC. In this case we have the additional 
ambiguity that the occupancy of the inert orbitals, absent in the Hamiltonian solved by CTQMC, lie between
2 and 3, for a total occupancy of 6 electrons. Therefore we have adjusted the chemical potential in
this Hamiltonian for two occupancies: i) 4 and ii) 3 electrons. In the first case we obtain a double occupancy of 
the $3d_{3z^2-r^2}$ and single occupancy of the $m= \pm 1$ orbitals, in agreement with Table \ref{tabu2} 
for large $U$. In case ii) we obtain one electron in each of the non-inert orbitals ($d_{3z^2-r^2}$, $d_{xz}$, $d_{xz}$),
resulting in the Co configuration with an additional $3d_{3z^2-r^2}$ hole, mentioned above.
Unfortunately our results are not enough to determine which of these two configurations is more probable.

\subsection*{Co}
As shown before \cite{dinap}, the configuration of Co corresponds to two holes occupying the $|m|=1$ orbitals and the 
remaining one is in one of the inert $|m|=2$ orbitals, forming a total spin $S=3/2$ according to Hund's rules. There is 
some admixture with the $S=1$ ground-state configuration of Ni described below.
The low temperature physics corresponds to an underscreened Kondo model, $S=3/2$ screened by two channels.
It is interesting that under an appropriate tetragonal crystal field, the orbital degeneracy of the $|m|=2$ orbitals is 
broken and at the same time the spin-orbit coupling leads to a physics similar to the $S=1/2$ two-channel Kondo model 
with non-Fermi liquid properties. This has been discussed in detail elsewhere \cite{dinap,dinap2,dinap3}. As stated 
above, Fe impurities might display similar physics.

\subsection*{Ni}
From the occupancies listed in Table \ref{tab2}, one realizes that Ni fluctuates between a configuration with two holes 
and total spin 1 in the degenerate orbitals with $|m|=1$ and one with one hole in one of these orbitals. A closer 
investigation of the effective model reported before indicates that the first configuration dominates and the system is 
in the Kondo regime \cite{nickel}. This is confirmed by our NRG calculations on the effective model $H_{\rm eff}$.
Thus the system can be considered as a realization of a two-channel fully screened $S=1$ Kondo model. The effective 
Hamiltonian including charge fluctuations is \cite{nickel}

\begin{eqnarray} \label{ham}
&H_{\rm eff}&=\sum_{M_{2}}(E_{2}+
D M_{2}^{2})|M_{2}\rangle \langle
M_{2}|+\sum_{\alpha M_{1}}E_1|\alpha M_{1}\rangle \langle \alpha M_{1}| + \nonumber\\
&&\sum_{\nu k\alpha \sigma }\epsilon _{\nu k}
c_{\nu k\alpha \sigma}^{\dagger }c_{\nu k\alpha \sigma }+\\
&&\sum_{M_{1}M_{2}}\sum_{\alpha\nu k\sigma }V_{\nu\alpha }\langle 1
M_{2}  |\frac{1}{2}\frac{1}{2}M_{1}\sigma \rangle 
( | M_{2}\rangle \langle \alpha M_{1}|c_{\nu k\alpha \sigma }+\mathrm{H.c.} ),  \nonumber 
\end{eqnarray}
where $E_i$ and  $M_i$ indicate the energies and the spin projections along the chain, chosen as the 
quantization axis, of states with $i=1, 2$ holes in the $3d$ shell of the Ni impurity, and 
$c_{\nu k\alpha \sigma}^{\dagger }$ creates a hole in the conduction band with symmetry $\alpha$ and spin 
$\sigma$ at the left or right of the impurity (denoted by the subscript $\nu$) with wave vector $k$.

\begin{figure}[ht!]
\includegraphics[width=0.70\columnwidth]{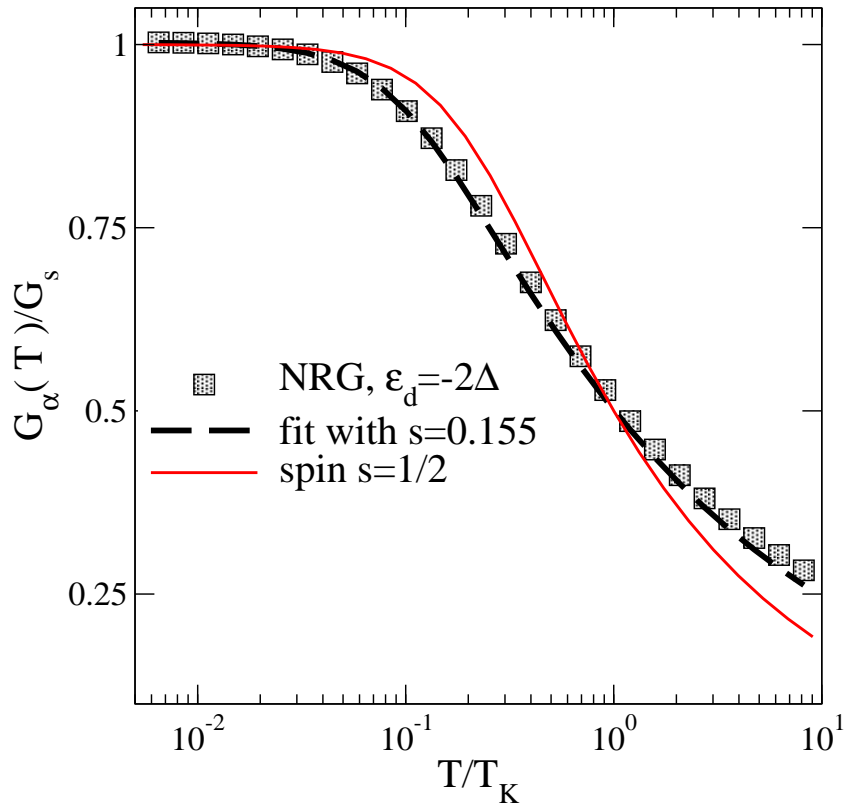}
\caption{(Color online)Squares: NRG data for $G_{\alpha}(T)=\sum_{\sigma}G_{\alpha\sigma}(T)$ 
in units of its maximum $G_s$ as a function of $T/T_K$ for $\epsilon_d=-2\Delta$,
where $\Delta = 0.115$ eV is half the resonant-level width \cite{nickel}.
Dashed black line: fitting of numerical data with expression
Eq.  (\ref{ley_empirica}) with $s=0.155$. Continuous red line: 
Eq.  (\ref{ley_empirica}) with $s=0.22$. }
\label{gs1}
\end{figure}

The ground state of this model is a Fermi liquid, as for the one-channel $S=1/2$ Kondo model.
The reader might ask if there are important differences in the properties of both models. One difference is that
well inside the Kondo regime, the conductance at zero temperature is two times larger in the 
two-channel $S=1$ case ($4 e^2/h$ for symmetric contacts) because two conduction channels and two spins
are contributing. Experimentally, asymmetry between the couplings of the impurity to the left 
and the right decreases the conductance, but this difference is likely observable in real experiments.
Another difference is that the {\em shape} of the conductance as a function of temperature is different in the
two cases. This is shown in Fig. \ref{gs1}, where the conductance is calculated for the parameters extracted in Ref. 
\cite{nickel}. For the one-channel $S=1/2$ Anderson model, it is well known that the empirical expression 
\begin{equation}
G(T)=\frac{G_{s}}{\left[ 1+\left( 2^{1/s}-1\right) \left( T/T_{K}\right)
^{2}\right] ^{s}},  \label{ley_empirica}
\end{equation}
where $s=0.22$ and $G_s$ is the conductance at temperature $T=0$, matches not only the experimental results but also 
NRG calculations \cite{gold1, G_E}. For the parameters that we extract for Ni in O-doped Au chains, we obtain that a 
similar empirical law holds but with $s=0.155$, yielding a less steep decrease in the intermediate temperature regime.

\subsection*{Cu}
Since Cu is similar to Au, it has an almost filled $3d$ shell, and therefore oxygen doping moves the spectral density 
of the orbitals with $|m|=1$ to the Fermi energy, introducing a partial emptying of these orbitals~\cite{dinap2}. We 
analyze the effect of the presence of oxygen impurities on the spin-state of the Cu impurity or its hole's 
symmetries, for several configurations, also changing the amount of O in the Au chain. In Table~\ref{Cu}  we introduce
the different cases studied and their corresponding Cu occupancy numbers (in electrons) obtained by integrating the 
minority band separated in the different symmetries, within the Cu-muffin-tin sphere. From the last configuration 
listed in Table~\ref{Cu}, it can be inferred that for this case a hole of the Au-O conduction bands with symmetry either $xz$ 
or $yz$ and spin either up or down can enter the full $3d$ shell of Cu or vice versa. Thus, this system can be 
described by an SU(4) Anderson model \cite{mina,joaq}. 

\begin{table*}[ht!]
\caption{Symmetry-dependent $d$-band minority spin fillings of the Cu atoms (in electrons) for the selected studied 
cases. The color coding of the schematic representation of the chains is the one presented in Fig.~\ref{3d-imp}}
\label{Cu}
\begin{indented}
\item[]\begin{tabular}{@{}cccc}
\br
Case &$n(3d_{3z^2-r^2}\downarrow)$& $n(3d_{xy,x^2-y^2}\downarrow)$ &  $n(3d_{xz,yz}\downarrow)$ \\
\mr
\includegraphics[width=0.4\columnwidth]{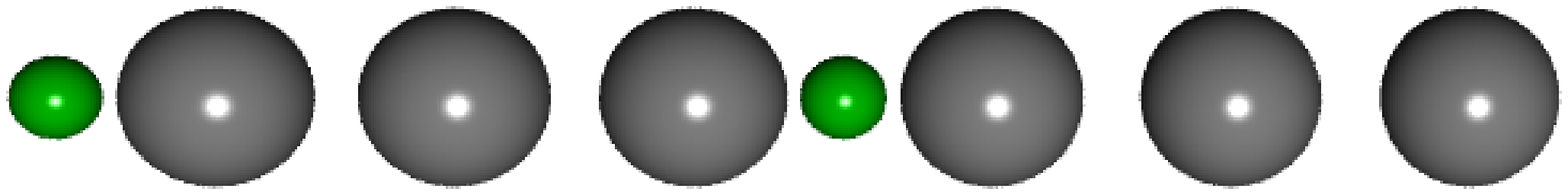}         & 0.83 & 1.89  & 1.87           \\
\includegraphics[width=0.4\columnwidth]{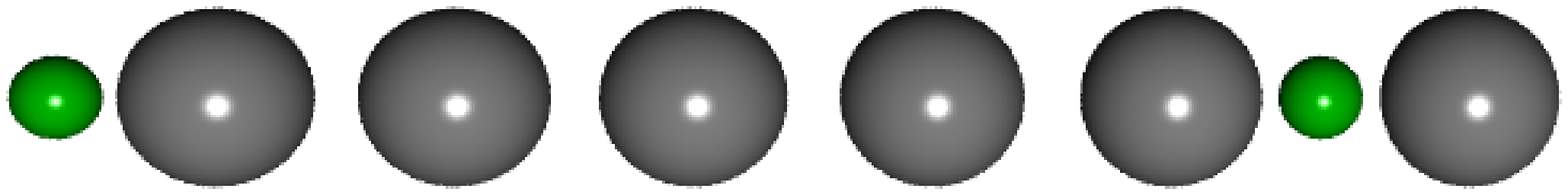}         & 0.91 & 1.89  & 1.64           \\
\includegraphics[width=0.4\columnwidth]{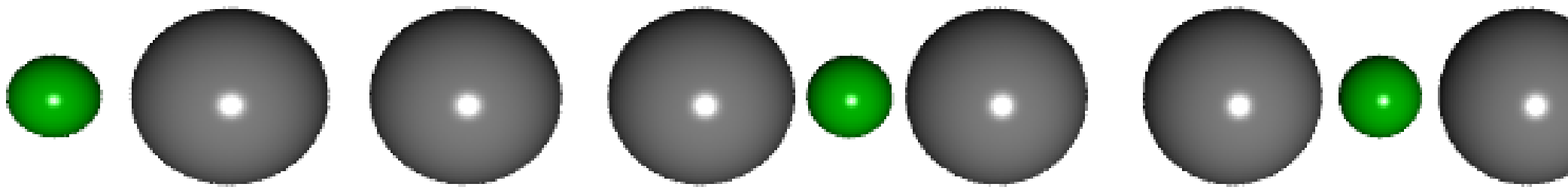}     & 0.90 & 1.89  & 1.65           \\
\includegraphics[width=0.4\columnwidth]{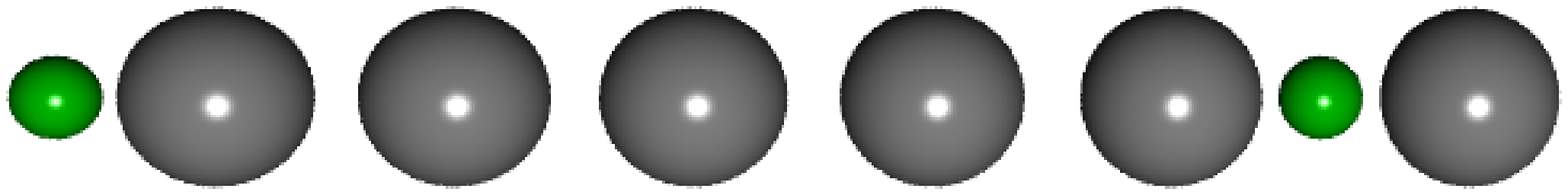}  & 0.83 & 1.79  & 1.36            \\
\br
\end{tabular}
\end{indented}
\end{table*}

\begin{figure}[h!]
\vspace{0.5cm}
\includegraphics[width=0.70\columnwidth]{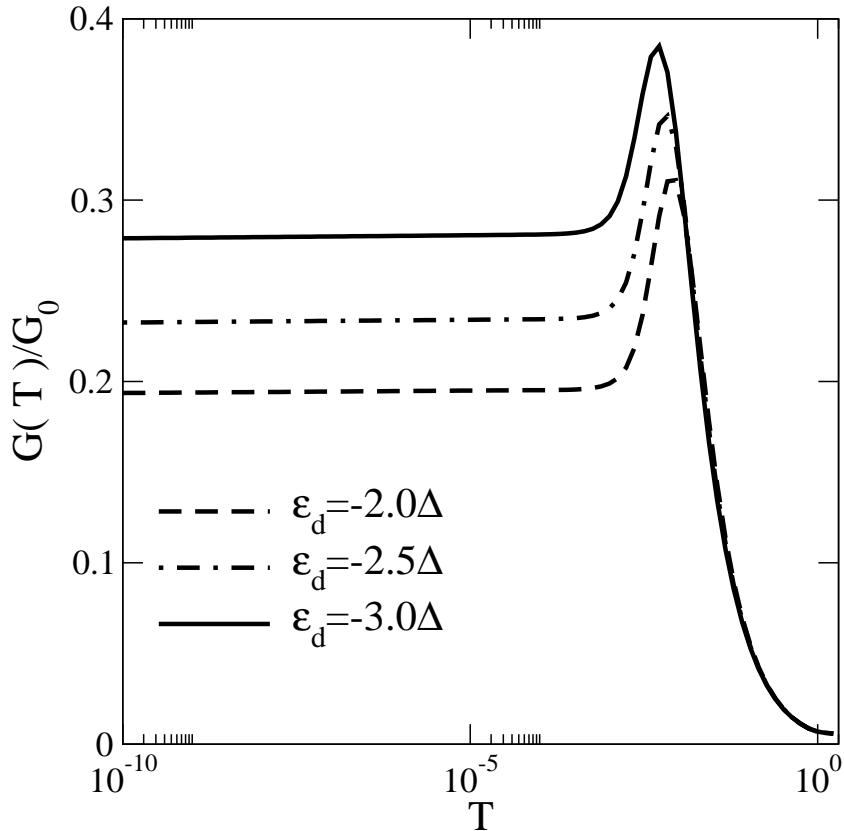}
\caption{(Color online) Total NRG conductance $G(T)$ for the 
SU(4) Anderson model as a function of temperature, 
for three values of $\epsilon_d$ ($-2 \Delta$, $-2.5 \Delta$ and $ -3\Delta$) with
hybridization $\Delta = 0.01$ in units of the half-bandwidth $D$.}
\label{gcu}
\end{figure}

In Fig. \ref{gcu} we show the NRG result for the total conductance $G(T)$ as a function of temperature
in units of $G_0= 2 e^2/h$ for three values of the on-site energy of the $xz$ and $yz$ orbitals
relative to the Fermi energy, $\epsilon_d$, and hybridization $\Delta = 0.01$ in units of the half-bandwidth $D$. 
All of them correspond to the intermediate valence regime with a total occupancy
of the $3d$ shell of 0.484 for $\epsilon_d=-3 \Delta$,  0.439 for $\epsilon_d=-2.5 \Delta$ and 0.398 for $\epsilon_d=-2 \Delta$, 
shared equally between the four
spin-orbitals. Note that the intermediate-valence regime in the present SU(4) case extends to larger values of 
$-\epsilon_d/\Delta$ than in the most usual SU(2) case.  
There is a clear maximum at temperatures of the order of $\epsilon_d$. At smaller temperatures, 
the conductance is rather flat, in contrast to other fully screened cases, such as those 
displayed in Fig. \ref{gs1}. 

At $T=0$ the conductance is determined by the Friedel sum rule. For constant
density of states and hybridization (as we have assumed for the NRG results), according to this rule,
the contribution of the conductance for each spin and channel with 
occupancy $n_{\alpha \sigma}$ is \cite{restor}

\begin{equation}
G_{\alpha \sigma}=\frac{e^2}{h} {\rm sin}^2( \pi n_{\alpha \sigma}),
\label{fsr}
\end{equation}
and $G= \sum_{\alpha \sigma}G_{\alpha \sigma}$.

The numerical results of Fig. \ref{gcu} are quite consistent with this rule. They lie above those
obtained using Eq. (\ref{fsr}) by near 1\%. This is likely due to numerical errors in the conductance.

\section{Summary and discussion}
\label{summ}

Using {\it ab initio} methods combined with continuous-time quantum Monte Carlo, we have studied the electronic
structure of systems with one substitutional $3d$ impurity in oxygen doped gold chains, along the whole $3d$ series,
searching for unusual behavior in the screening of the impurity spin.
The effect of oxygen doping is to bring $5d_{xz}$ and $5d_{yz}$ orbitals to the Fermi energy, which together
with the $5d_{3z^2-r^2}$ ones, constitute three screening channels for the impurity spin.

We left Zn out because one expects that it has a full $3d$ shell and no magnetism. At the other
end of the $3d$ series, we find that Sc behaves as an isolated magnetic impurity with spin $S=1/2$, because it
has one electron occupying either the $3d_{x^2-y^2}$ or the $3d_{xy}$ orbitals, and they have a negligible hybridization
with the orbitals of the rest of the chain. For this reason we call them inert.

The system with Ni and Cu impurities behave as a Fermi liquid, obeying Friedel sum rule. 
In the first case, the low-energy physics
can be described by a fully compensated $S=1$, two-channel Kondo model. In this case, we have calculated the 
conductance through the system using NRG, pointing out quantitative differences with the most usual $S=1/2$, 
one-channel case. For Cu impurities, the appropriate low-energy model is the SU(4) Anderson impurity one, in
the intermediate-valence regime. In this case, the conductance also shows differences with more usual cases.

The remaining systems, taking out some peculiarities, are expected to behave as singular Fermi liquids, 
like underscreened Kondo model. This is due to the presence of partially filled inert orbitals which 
tend to couple into large spins due to the Hund rules, and cannot be screened by the conduction electrons.
The conductance as a function of temperature for several underscreened Kondo models is presented 
in the Supplementary material of Ref. \cite{parks2}.
Particularly complex are the case of Cr in which different configurations seems mixed and Fe, for which our results
are inconclusive with respect of the ground state configuration. One possibility for Fe is that 
it has the same configuration of Co with one additional $3d_{x^2-y^2}$ hole. The spin of this hole is
expected to be screened in a first-stage Kondo effect leaving a low-energy physics similar to that of Co.
As shown before for the latter \cite{dinap,dinap2,dinap3}, under an appropriate tetragonal crystal field, 
which can be realized connecting the system with leads with a square cross section, the 
spin-orbit coupling leads to an effective spin 1/2 overscreened by two degenerate channels with $xz$ and $yz$ 
symmetries, leading to non-Fermi-liquid behavior.

The results using CTQMC have been done using a restricted basis set. This has the advantage of accelerating
the time in the calculations and the interactions take a simpler form than those of the complete basis set 
\cite{kroll}. The main drawback however is then the Hund interaction with the spin  of the inert orbitals is 
neglected. We believe that this does not alter the resulting occupancies of Cr and Fe,
because the resulting spin of the remaining orbitals is the maximum possible, taking maximum advantage of 
the Hund interaction.

Finally we comment on the effect of symmetry-breaking perturbations, like distortions or defects in the chains.
This effect breaks the channel degeneracy and changes the physics in those cases in which the impurity has partial
degenerate orbitals which are not inert. For example in the case of Ni, breaking of the degeneracy of the 
$xz$ and $yz$ orbitals would change the $S=1$ two-channel Kondo physics, in a two-stage Kondo model with two 
characteristic temperatures, each one corresponding to each channel. The ground-sate however continues to be a 
Fermi liquid.  In the case of Co, the partial screening of the $S=3/2$ spin by two channels also would occur 
in two stages, but the ground state still corresponds to a $S=1/2$ singular Fermi liquid. The breaking of symmetry 
is more dramatic for Co in tetragonal symmetry, because the symmetry breaking renders the effective $S=1/2$ 
overscreened model with non-Fermi liquid behavior to a usual Fermi liquid below a characteristic energy scale.
A previous estimate indicates that this scale is below $T_K$ (and the non-Fermi liquid features are observable)
if the splitting between $xz$ and $yz$ orbitals is less than 100 meV \cite{dinap}.

\section*{Acknowledgments}

This work was sponsored by PIP 112-201101-00832, 112-201101-0160 and 112-201201-00069 of CONICET, PICT 2013-1045 and 
PICT-2014-1555 of the ANPCyT.

\end{document}